\documentclass[11pt]{article}
\usepackage{mathpazo}

\usepackage[normalem]{ulem}
\usepackage{amsmath,bm}
\usepackage{graphicx,pdflscape}
\usepackage{color}
\usepackage[toc,page]{appendix}
\usepackage{simpler-wick}

\usepackage{color,subfigure}

\usepackage{hyperref}
\hypersetup{
    colorlinks,
    citecolor=red,
    filecolor=cyan,
    linkcolor=blue,
   urlcolor=magenta
 }

\newcommand{\disp}[1]{Eq.~(\ref{#1})}

\newcommand{\figdisp}[1]{Fig.~(\ref{#1})}

\newcommand{\lessim} {\ {\raise-.5ex\hbox{$\buildrel<\over\sim$}}\ }

\newcommand{\gssim}{\ {\raise-.5ex\hbox{$\buildrel>\over\sim$}}\ }

\newcommand{\nn}{\nonumber}

\newcommand{\w}{\omega_0}

\renewcommand{\emph}{\textit}

\newcommand{\iden}{ {\bf 1}}

\usepackage{hyperref}
\usepackage{color}

\newcommand{\half}{\frac{1}{2}}
\newcommand{\beq}{\begin{eqnarray}}
\newcommand{\eeq}{\end{eqnarray}}
\newcommand{\barray}{\begin{eqnarray}}
\newcommand{\earray}{\end{eqnarray}}
\newcommand{\kp}{k_{\small ||}}

\renewcommand{\v}[1]{| #1 \rangle}
\newcommand{\h}[1]{\widehat{#1}}
\begin{document}

\title{  The Toeplitz matrix $e^{- \kappa |i-j|}$ and its application to a layered electron gas}

\author{Onuttom Narayan and  B Sriram Shastry  \\
\small \em Physics Department, University of California, Santa Cruz, CA, 95064 \\}
\date{\today}

\maketitle

\abstract{We present an explicit solution of the eigen-spectrum of
the Toeplitz matrix $C_{ij}= e^{- \kappa |i-j|}$ with $0\leq i,j
\leq N,$ and extend it to a combination of a Toeplitz matrix and a
Hankel matrix.  The solution is found by   elementary means that
bypass the Wiener-Hopf technique usually used for this class of
problems. It  rests on the observation that the inverse of $C_{ij}$
is effectively a nearest neighbor hopping model with specific
onsite energies, which can in turn be diagonalized easily. We apply
this result to find analytically the plasma modes of a layered 
assembly of 2-dimensional electron gas. We find  a sum rule relating the geometric mean of 
the frequencies of the plasma modes to the determinant of this Toeplitz matrix.}


\newpage
\noindent {\bf \S Introduction:} 

In the course of our study of layered electronic systems initiated
in\cite{Visscher}, we came across an interesting Toeplitz matrix
\beq
C_{ij}=  e^{- \kappa |i-j|}, \; \hspace{.2in} 0\leq i,j \leq N \label{Cij}.
\eeq
Eigenfunctions and eigenvalues of Toeplitz matrices are usually
found by the Wiener-Hopf technique\cite{Its,Widom}.  The special
case of Eq.(\ref{Cij}) is called a Kac-Murdock-Szego matrix\cite{KMS},
which can be solved\cite{For1,Horn,Trench,Poddubny} by noting that its inverse
is a simple tridiagonal matrix, whose eigenfunctions and eigenvalues
can easily be found. 
An aspect of interest is that the inverse matrix is a tridiagonal
matrix, i.e. a nearest neighbor hopping model with a specific on-site
and boundary energy. This type of  matrix  arises in a large number
of problems in condensed matter physics, and therefore the relationships
found here may thus be of broad interest.

The method of finding the eigenfunctions and eigenvalues of the
matrix $C_{ij}$ by constructing its inverse can be generalized to
a matrix that is the combination of a Hankel matrix and a Toeplitz matrix:
\beq
M_{ij} = a e^{- \kappa |i - j|} + b e^{\kappa |i - j|}
+ c \left[e^{-\kappa(i + j)} + e^{- \kappa (2 N - i - j)}\right]
\label{Mij}
\eeq
with arbitrary $a, b$ and $c.$ This generalization is the subject
of this paper.

The particular combination of Hankel and Toeplitz matrices given
above is appropriate for solving the plasma modes of a layered
assembly of 2-dimensional electron gas. The plasma mode frequencies
have been found numerically for this system by a method different
from ours\cite{Jain,SDS,Fetter}, and has in fact been studied
experimentally using Raman scattering\cite{Pinczuk}.  Applying our
method to this problem yields a simple expression for the eigenfunctions
associated with these plasma modes, as well as a sum rule relating
the frequencies of the (N+1) branches of the plasma frequency as
functions of the parallel component of the photon wave vector.  The
density of states of eigenvalues is also of interest
experimentally\cite{Bozovic} and evaluated analytically here.

In the rest of this paper, we first review the details of the
inversion of C, and two  matrices closely related to it, followed
by a calculation of the eigenspectrum.  We then generalize the
method to calculate the eigenspectrum of the matrix $M$ in
Eq.(\ref{Mij}). Finally, we apply the result to the layered electron
gas.


  \vspace{.25 in}
\noindent {\bf \S The Inversion of $C_{ij}$}

For $0\leq j \leq N$ we denote the basis column vector $\hat{e}_j$ (with1 at the $j^{th}$ row and 0 elsewhere) as  $\v{j}$, and write an operator $\h{C}$ such that $\h{C} \v{j}= \sum_{l=0}^N C_{lj} \v{l} $. Let us also denote $\w=e^{\kappa}$. We  decompose $\h{C}$ into right ($\h{R}$) and left ($\h{L}$) moving parts as 
\beq
\h{C}+ \iden= \h{R}+\h{L},
\eeq
where $\iden$ is the identity operator.  The operators $\h{R}$ and $\h{L}$  are defined  by their action on the basis states
\beq
\h{R} \v{j} = \sum_{l=j}^N e^{- \kappa (l-j)} \v{l}= \v{r_j}\equiv  \sum_{l=j}^N \frac{\w^j}{\w^l}\;\;\v{l} \\
\h{L} \v{j} = \sum_{l=0}^j e^{-\kappa (j-l)}\v{l}= \v{l_j}\equiv \sum_{l=0}^j \frac{\w^l}{\w^j} \;\; \v{l} \label{RL}.
\eeq
Note that  the boundary vectors are given by
\beq
\v{l_0}=\v{0}, \mbox{ and} \;\; \v{r_N}=\v{N}. \label{bct}
\eeq

For the next steps it is useful to note four recursion relations between  the basis vectors and their domains
\beq
&&\v{l_{j+1}}= \frac{1}{\w} \v{l_{j}}+\v{j+1},\;\; \mbox{for} \;\; 0\leq j \leq N-1 \label{g1} \\
&&\v{l_{j-1}}= {\w} \v{l_{j}}- \w \v{j},\;\;\;\; \mbox{for} \;\; 1\leq j \leq N \label{g2}\\
&&\v{r_{j+1}}= \w \v{r_j} - \w \v{j}, \;\;\;\; \mbox{for}\;\; 0 \leq j \leq N-1 \label{g3} \\
&&\v{r_{j-1}}= \v{j-1} + \frac{1}{\w} \v{r_j} \;\;\;\; \mbox{for}\;\; 1 \leq j \leq N. \label{g4}
\eeq

 Let us calculate the action of $\h{C}$ on the states. Consider first the {\em interior terms} $1\leq j \leq N-1$:
 \beq
 (\h{C}+\iden)\v{j+1}&=& \v{r_{j+1}}+\v{l_{j+1}} \nn\\
 &=& \w \v{r_j}+ \frac{1}{\w} \v{l_j}+\v{j+1}-\w \v{j}, \label{eq10}
 \eeq
 using Eqs.~(\ref{g1},\ref{g3}). Similarly
 \beq
 (\h{C}+\iden)\v{j-1}&=& \v{r_{j-1}}+\v{l_{j-1}} \nn\\
 &=& \frac{1}{\w} \v{r_j}+ \w \v{l_j} - \w \v{j}+ \v{j-1} \label{eq11} 
 \eeq
 using Eqs.~(\ref{g2},\ref{g4}). Adding \disp{eq10} and \disp{eq11} and rearranging we find
 \beq
 \h{C}(\v{j+1}+\v{j-1})= (\w+\frac{1}{\w}) \h{C} \v{j} - (\w-\frac{1}{\w}) \v{j}. 
 \eeq
 Multiplying through by the inverse operator $ \h{C}^{-1}$ we find
 \beq
 \h{C}^{-1}\v{j}= \coth{\kappa} \v{j}- \frac{1}{2 \sinh{\kappa}} (\v{j+1}+\v{j-1}). \label{body}
 \eeq  
To determine the action of $ \h{C}^{-1}$ on the boundary term $j=0$ we note
\beq
(\iden+\h{C})\v{1}&=& \v{l_1}+\v{r_1}=\v{1}+\frac{1}{\w} \v{0}+\w \v{r_0}-\w \v{0},\nn\\
\h{C} \v{1}&=&\w \v{r_0}- (\w-\frac{1}{\w}) \v{0}
\eeq
where we used \disp{bct}, \disp{g1} and \disp{g3}. Now observe that on using \disp{bct}
\beq \h{C}  \v{0}= \v{r_0},
\eeq
 and hence we may write
\beq
\h{C} \v{1}= \w \h{C}\v{0} - (\w-\frac{1}{\w}) \v{0},
\eeq
or taking the inverse,
\beq
\h{C}^{-1} \v{0}= \frac{e^\kappa}{2 \sinh{\kappa}} \v{0}- \frac{1}{2 \sinh{\kappa}} \v{1}. \label{left}
\eeq
To determine the action of $ \h{C}^{-1}$ on the boundary term $j=N$ we note that
\beq
\h{C}\v{N-1}&=& \w \v{l_N}-\w \v{N} +\frac{1}{\w} \v{N}
\eeq
where \disp{bct} and \disp{g4} have been used. We further use 
\beq
\h{C}\v{N}= \v{l_N},
\eeq
so that 
\beq
\h{C}\v{N-1}&=& \w \h{C}\v{N}-\w \v{N} +\frac{1}{\w} \v{N}.
\eeq
Upon inversion we get
\beq
\h{C}^{-1} \v{N}=\frac{e^\kappa}{2 \sinh{\kappa}} \v{N}- \frac{1}{2 \sinh{\kappa}} \v{N-1}. \label{right} 
\eeq
Combining \disp{left}, \disp{right} and \disp{body} we write the inverse matrix in the form of a tight-binding Hamiltonian
\beq
&&\h{C}^{-1}= \sum_{j=0}^N \varepsilon(j) \v{j}\langle j| -\tau \sum_{j=0}^{N-1} \left\{ \v{j}\langle j+1|+ \v{j+1}\langle j| \right\}, \nn \\
&& \tau=  \frac{1}{2 \sinh{\kappa}}\nn \\
&&\varepsilon(1)=\varepsilon(2)=\cdots=\varepsilon(N-1)= \coth{\kappa}, \nn \\
&&\varepsilon(0)=\varepsilon(N)=\frac{e^\kappa}{2 \sinh{\kappa}};\;\;  \label{Inv}
\eeq

  \vspace{.25 in}
\noindent {\bf \S Inverses of $\h{R}$ and $\h{L}$}

It is interesting to note the inverses 
\beq
\h{R}^{-1}= \iden - e^{-\kappa} \sum_{j=0}^{N-1} \v{j+1} \langle j | \label{Rinv}
\eeq
i.e. the identity  minus a right shift operator, and 
\beq
\h{L}^{-1}= \iden - e^{-\kappa} \sum_{j=0}^{N-1} \v{j} \langle j+1 | \label{Linv}
\eeq
i.e. the identity  minus a left shift operator. The proof uses a similar idea as before. For \disp{Rinv} with $0\leq j \leq N-1$, we use \disp{g3} to write
\beq
\h{R} \v{j+1}=  \w \h{R} \v{j}- \w \v{j}, \; \mbox{or} \;\; \h{R}^{-1} \v{j}= \v{j}- \frac{1}{\w} \v{j+1},
\eeq
and for the boundary term use $\h{R}^{-1} \v{N} = \v{N}$. For \disp{Linv} with $0\leq j \leq N-1$, we use \disp{g1} to write
\beq
\h{L}\v{j+1}= \frac{1}{\w} \h{L} \v{j} + \v{j+1}, \;\;\mbox{or}\;\; \h{L}^{-1} \v{j+1}= \v{j+1}- \frac{1}{\w} \v{j},
\eeq  
and for the boundary term $\h{L}^{-1}\v{0}=\v{0}$. Together these result in \disp{Rinv} and \disp{Linv}.

  \vspace{.25 in}
\noindent {\bf \S Diagonalizing  of $\h{C}$}

It is actually  easier to diagonalize $\h{C}^{-1}$ in \disp{Inv}. We try the wave function
\beq \v{\Psi(q)}= \sum_{j=0}^N \cos( q j - \Phi(q)) \v{j} \label{psi}
\eeq
such that
\beq
\h{C}^{-1} \v{\Psi(q)} = \Lambda^{-1} \v{\Psi(q)}.
\eeq
Here $q$ and $\Phi(q)$ as well as the eigenvalue $\Lambda$ are to be determined.
The interior terms $1\leq j \leq N-1$ are satisfied by this wavefunction provided
\beq
\Lambda^{-1}= \coth \kappa - \frac{\cos q}{\sinh \kappa},
\eeq
and the amplitude at $j=0$ requires the condition
\beq
(\Lambda^{-1}- \varepsilon(0)) \cos \Phi= -\tau \cos(q -\Phi),
\eeq
or simplifying further we find the phase shift determined by
\beq
\Phi(q)= \mbox{arccot} \left\{ \frac{\sin q}{\cos q  - e^{-\kappa}} \right\}.
\eeq
The phase shift   $\Phi(q)$ varies  continuously with $q$ in the interval $0\leq q\leq \pi$,  decreasing monotonically from $\pi/2$ to $-\pi/2$. It is thus a convenient parameterization for finding all the eigenvalues. The amplitude at $j=N$ is satisfied if 
\beq
(\Lambda^{-1}- \varepsilon(N)) \cos (q N- \Phi(q))= -\tau \cos(q (N-1) -\Phi(q)),
\eeq
or simplifying further
\beq
\sin(q N - 2 \Phi(q))&=&0. 
\eeq
Alternatively, we can observe that the eigenfunctions must be odd or even functions of the index $j$ measured from the midpoint of $j = N/2$ (this is true even if $N$ is odd), so that either $\sin (q N/2 - \Phi(q))$ or
$\cos (q N/2 - \Phi(q))$ is zero for each eigenfunction. The product of the two expressions, and therefore $\sin(q N - 2 \Phi(q))$ must therefore be zero for every eigenfunction.
It is straightforward to  verify that the N values  $ \nu=0,1,\ldots N$ yield the $N+1$ distinct eigenvalues 
\beq
\Lambda(q_\nu,\kappa)&=& \frac{\sinh \kappa}{\cosh \kappa - \cos q_\nu}, \label{ev} \;\;\mbox{with} \\
 q_\nu N&=&  \nu \pi  + 2 \Phi(q_\nu). \label{qs}
\eeq 
We will usually denote $\Lambda(q_\nu,\kappa)$ as $\Lambda(q_\nu)$.
At finite $N$ the values $q=0$ and $q= \pi$ are excluded since  for these the wavefunction $\v{\Psi(q)}$ vanishes identically, formally these correspond to $\nu=-1$ and $\nu=N+1$ respectively. Also we note that in the limit $\kappa\to + \infty$, the phase shift $\Phi(q)= \pi/2-q$ and hence $q_\nu= \frac{\nu+1}{N+2} \pi$.

The  matrices $\h{R}$ and $\h{L}$  act as raising or lowering operators and do not have the usual eigenfunctions, however it is easy to construct their generalized eigenfunctions.

  \vspace{.25 in}
\noindent {\bf \S Density of states}

For large N it is useful to employ the density of states of the exact eigenvalues, these can be found straightforwardly.
We note the identity
\beq
\frac{d \Phi(q)}{d q}=- \half (1+ \Lambda (q)),
\eeq
so that we can write the difference in successive solutions from  \disp{qs} in the form
\beq
\pi \Delta \nu = N \Delta q_\nu - 2 \Delta \Phi(q_\nu)= \Delta q_\nu ( N+1 + \Lambda(q_\nu))
\eeq
so that 
\beq
\sum_{\nu=0}^N  \to \int_{q_0}^{q_N} \frac{ d q}{ \pi} \left\{ N + 1 + \Lambda(q) \right\}.
\eeq
From \disp{qs} 
\beq
\frac{d q}{d \Lambda}= - \frac{\sinh \kappa}{\Lambda^2 \left[ 1- ( \cosh \kappa - \frac{ \sinh \kappa}{\Lambda})^2\right]^\half}
\eeq
and hence we can convert  a sum over solutions to an integral over eigenvalues with a density of states  
\beq
\sum_\nu \to \frac{1}{\pi} \int_{\Lambda_<}^{\Lambda_>} \frac{d \Lambda}{\Lambda^2} \frac{ \left\{ N + 1 + \Lambda \right\} \sinh \kappa}{\Lambda^2 \left[ 1- ( \cosh \kappa - \frac{ \sinh \kappa}{\Lambda})^2\right]^\half} \label{DOS}
\eeq
where
\beq
\Lambda_< = \frac{\sinh \kappa}{\cosh \kappa +1 }, \;\;  \Lambda_>= \frac{\sinh \kappa}{\cosh \kappa -1 }.
\eeq

  \vspace{.25 in}
\noindent {\bf \S Szeg\H{o}'s theorem for the determinant of $C_{ij}$}

It is interesting to compute the determinant of $C$.  For the matrix \disp{Cij} we are in the happy position of being able to do so exactly by using Gauss's method of triangulation, leading to
\beq
 ||C||= det(C)= (1- e^{-2 \kappa})^N. \label{det}
\eeq
The proof is elementary. An alternative approach exploits the tridiagonal nature of $C^{-1}.$ If one defines $A_j$ to be the $j\times j$ submatrix of $(2\sinh\kappa) C^{-1}$ 
that ends at the bottom right corner of $C^{-1},$ it is easy to verify that $||A_{j+1}|| = V_j ||A_j|| - ||A_{j-1}||$ for $j = 1,2\ldots N$ with the boundary condition $||A_0|| = 1$ and $||A_1|| = e^\kappa,$ and 
$V_j = e^\kappa \delta_{j,N} + (1 - \delta_{j, N}) 2\cosh\kappa.$ With the boundary condition, the solution to the recurrence relation is $||A_j|| = e^{j\kappa}$ for $0 \leq j \leq N,$ and so
$||A_{N+1}|| = e^{(N+1)\kappa} (1 - e^{-2\kappa}).$ Therefore $||C|| = (2 \sinh\kappa)^{N+1}/[e^{(N+1)\kappa} (1 - e^{-2\kappa})] = (1 - e^{-2\kappa})^N.$

We can also calculate the determinant from the strong theorem of Szeg\H{o} \cite{Szego}, which is guaranteed to give the two leading terms in the limit of large N.
Specifically the theorem says that when  the (N+1)x(N+1) Toeplitz matrix $C$  is generated by a density $\phi(\theta)$ through a Fourier series, i.e. 
\beq
 C(i-j)= \int_{-\pi}^\pi \; \frac{d \theta}{2 \pi} \; e^{- i \theta (i-j)} \phi(\theta) \label{phi}
\eeq
and further if 
 \beq
\log \phi(\theta) = \sum_{l=-\infty}^\infty e^{i l \theta} \nu_l \label{logphi}
\eeq
then the determinant  for large $N$ is given by 
\beq
||C||=\exp\{(N+1) \nu_0 + \sum_{l=1}^\infty l |\nu_l|^2 + o(N)\} \label{strong}.
\eeq
In the present case of \disp{Cij} it is readily seen that
\beq
\phi(\theta)& =& \frac{\sinh \kappa}{\cosh \kappa-\cos(\theta)}, \;\;\mbox{and} \\
\nu_l&=&\delta_{l,o} \left(\log 2 \sinh \kappa - \kappa \right)+(1-\delta_{l,o}) \frac{e^{- \kappa |l|}}{|l|}.
\eeq
Substituting into \disp{strong} and carrying out the summation over $l$ we see that Szeg\H{o}'s theorem gives
\beq
||C||&=& \exp\{(N+1) \left[ \log (1-e^{- 2 \kappa}) \right] -\left[  \log(1-e^{- 2 \kappa}) \right]+ o(N)\} \nn \\
&=& \exp\{ N \left[ \log (1-e^{- 2 \kappa}) \right] + o(N)\} \label{res-szego}.
\eeq
Comparing with \disp{det} we see that the above expression is exact
if we drop the $o(N)$ correction terms altogether. We can also
calculate the determinant using the exact eigenvalues $\Lambda$
given in \disp{qs} and employing the Euler-Mclaurin formula. The
first two terms are  the same as in \disp{res-szego}. The rather
unexpected vanishing of the $o(N)$ correction  term, as explained
to us by Prof. Ehrhardt, is the consequence of the following general
result\cite{WidT}: define $\varphi(z) = \sum_{k=-\infty}^\infty
C_{k + j, j} z^k.$ If $1/\varphi(z)$  has a Laurent series in which
all $\sim z^k$ terms vanish for $k > m $ or $k < -m,$ then the
determinant of $C$ is of the form $G^N E$ for $N+1 \geq m,$ for
some constants $E$ and $G.$ (The constants can be defined as $G =
||C_{N=m}||/||C_{N=m-1}||$ and $E = ||C_{N=m}||/G^m.$) In the case
at hand, $m=1,$ and the determinant grows exponentially with $N$
over the entire range of $N.$

  \vspace{.25 in}

  \vspace{.25 in}

\noindent {\bf \S Generalization to combined Toeplitz Hankel matrices}

Toeplitz matrices are closely related to Hankel matrices: the
elements $H_{ij}$ of a Hankel matrix $H$ only depend on $i+j.$ It
is clear that any Hankel matrix is related to some Toeplitz matrix
through reflection about the midpoint: $i \rightarrow N - i$ or $j
\rightarrow N - j.$ In particular, the matrix
\beq
\tilde H_{ij}=  e^{- \kappa |i+j - N|}, \; \hspace{.2in} 0\leq i,j \leq N \label{Hij}.
\eeq
is a reflection of the Toeplitz matrix $C_{ij}$ which we have
analyzed. Since $\tilde H = R C,$ where $R$ is the reflection
operator, any eigenvector of $C$ satisfies $\tilde H|\psi\rangle =
R C |\psi\rangle = \lambda R |\psi\rangle.$ Since, as we have
remarked earlier, the eigenvectors of $C$ are even or odd under
reflection about the midpoint, $\tilde H |\psi\rangle = (-1)^P
\lambda |\psi\rangle,$ where $P$ is the parity of the eigenvector.

A related Hankel matrix, $H_{ij} = \exp[-\kappa(i + j)],$ for which there
is no cusp on the diagonal, is even simpler to solve. It is easy
to verify that any vector $|\psi\rangle$ that satisfies $\sum_j
\exp[- j \kappa] \psi_j = 0$ is a null vector of $H.$ Thus the null
space of $H$ is $N$-dimensional, and the $N+1$'th eigenvector must
be the vector that is orthogonal to this subspace: $\psi_j = \exp[-
j \kappa]$  (unnormalized), with eigenvalue $\sum\exp[- 2 j \kappa].$

We now consider the problem of finding the eigenvalues of a combination of 
Hankel and Toeplitz matrices:
\beq
M_{ij} = a \exp\left[- \kappa |i - j|\right] + b \exp\left[\kappa |i - j|\right] 
+ c \left\{\exp\left[-\kappa(i + j)\right] + \exp\left[- \kappa (2 N - i - j)\right]\right\}
\eeq
with the restriction $a \neq b.$ Each of the four parts of this
matrix can be solved (in our discussion of the matrix $C,$ there
was no restriction that $\kappa$ had to be positive), but they are
non-commuting.

We define the tridiagonal matrix $T$ 
\beq
T = \frac{1}{2 (a - b) \sinh\kappa} 
\begin{pmatrix} 
	e^\kappa & -1 & 0 \ldots\\
	-1 & 2 \cosh\kappa & -1 \ldots\\
	\vdots &  & \ddots
\end{pmatrix}
\eeq
which is the same tridiagonal matrix we used earlier, except for the factor of $a-b$ in the denominator.
Then it is possible to verify that 
\beq
T M = I + 
\begin{pmatrix}
	\alpha_0 & \alpha_1 & \ldots \\
	0        &     0    & \ldots\\
	\vdots   & \vdots  & \ddots\\
	\alpha_N & \alpha_{N-1} & \ldots
\end{pmatrix}
\label{srulemod}
\eeq
i.e. the matrix $T$ is the inverse of $M$ except for boundary effects. Explicitly, the elements
of the boundary rows are 
\beq
\alpha_i = \frac{1}{a-b}(b \exp[\kappa i] + c \exp[-\kappa i]).
\label{alpha}
\eeq
The actual inverse
of $M$ is then 
\beq
M^{-1} = T + 
\begin{pmatrix}
        x_0 & x_1& \ldots \\
        0        &     0    & \ldots\\
        \vdots   & \vdots  & \ddots\\
        x_N& x_{N-1} & \ldots
\end{pmatrix}
\label{Minv}
\eeq
where, taking advantage of the symmetry properties of $M,$ the
condition to be satisfied by the $x_i$'s is
\beq
M 
\begin{pmatrix} x_0\\x_1\\\vdots\\ x_N \end{pmatrix}
	= - 
\begin{pmatrix} \alpha_0\\\alpha_1\\\vdots\\ \alpha_N \end{pmatrix}.
\eeq
This has the solution 
\beq 
\begin{pmatrix} x_0\\x_1\\\vdots\\ x_N \end{pmatrix}
	= -
	T \begin{pmatrix} \alpha_0\\\alpha_1\\\vdots\\ \alpha_N \end{pmatrix}
	-\begin{pmatrix} \sum \alpha_i x_i \\ 0 \\ \vdots \\ \sum \alpha_{N-i} x_i\end{pmatrix}.
\eeq
Substituting Eq.(\ref{alpha}) in the first term on the right hand side, all the elements of $T\cdot\alpha$ except the first
and last ones are zero. Therefore, $x_1, x_2 \ldots x_{N-1} =0$ and we are left with the coupled equations
\begin{equation}
	(a-b) \begin{pmatrix} x_0\\x_N\end{pmatrix} = 
		-\frac{1}{a-b} \begin{pmatrix} c\\ b e^{N\kappa}\end{pmatrix}
		-\begin{pmatrix} x_0 & x_N\\ x_N & x_0\end{pmatrix}
		\begin{pmatrix} b + c \\ b e^{N\kappa} + c e^{-N\kappa} \end{pmatrix}
\end{equation}
which has the solution 
\begin{eqnarray}
	x_0 &=& \frac{ac - bc + c^2 - b^2 e^{2\kappa N}}{(a - b) [(a + c)^2 - (c e^{-N \kappa} + b e^{N\kappa})^2}\nonumber\\
	x_N &=& \frac{c^2 e^{-N\kappa} - ab e^{N\kappa}}{(a - b) (a^2 + 2 (a - b) c - b^2 e^{2 N \kappa} + c^2 (1 - e^{-2 N\kappa})}.
\end{eqnarray}

Once we have obtained $M^{-1}$ in the form of Eq.(\ref{Minv}), it
is easy to see that the eigenvectors can be written with elements
$\psi_q(j)  = \cos[q (j - N/2)]$ or $\psi_q(j) = \sin[q (j - N/2)].$
The eigenvalues are related to $q$ through
\begin{equation}
	\Lambda^{-1} = \frac{1}{a - b} \left[\coth\kappa - \frac{\cos q}{\sinh\kappa}\right] .
\end{equation}
The boundary conditions for the even and odd eigenvectors
\begin{eqnarray}
	(x_0 + x_N) \cos q N/2 + \frac{e^{-\kappa}}{2(a - b)\sinh\kappa} [\cos q N/2 + \cos q (N/2 + 1)] &=& 0\nonumber\\
	(x_0 - x_N) \sin q N/2 + \frac{e^{-\kappa}}{2(a - b)\sinh\kappa} [\sin q N/2 + \sin q (N/2 + 1)] &=& 0\nonumber\\
\end{eqnarray}
respectively determine the allowed values of $q.$

From Eq.(\ref{srulemod}), it is easy to see that 
\beq
det(M) = (1 + \alpha_0)^2/det(T)  = (a - b)^{N-1} (a + c)^2 ((1 - e^{-2\kappa})^N.
\eeq

\noindent {\bf \S 2-d Plasmon spectrum}
 
As mentioned in the introduction, the Toeplitz  matrix $C_{ij}$ arises in the context of plasmons in multilayer systems, a system that has been studied extensively earlier. The original systems studied in the work of Olego, Pinczuk, Gossard and Wiegmann \cite{Pinczuk,Pinczuk-2} consists of alternating layers of insulating $GaAs$ and conducting $(Al_xGa_{1-x})As$. Here the conducting planes are coupled by the Coulomb interaction only, i.e. one ignores the direct hopping of electrons between layers \cite{Visscher}. Recent advances in materials allows a vast range of composite materials, generalizing this initial system \cite{Yan,DaSilva,Volodin,Okamoto}. To understand plasmons in these systems, one needs to understand the dielectric function of layered systems \cite{Fetter, SDS,Jain}, where the plasmon is a pole of a charge response function, probed by either a charged particle surface scattering,  or as in the case of \cite{Pinczuk, Pinczuk-2,Jain} by  photons using  Raman scattering. Within the widely  used random phase approximation for these systems, the plasmon is found as the eigen-solution of a homogeneous Fredholm equation\cite{Jain}  satisfied by $\delta\rho(l)$,  the induced charge density on layer $l$  due to a small excess external charge:
\beq
\delta \rho(l) = D_0(\kp,\omega) V(\kp)  \sum_{m=0}^N e^{- \kp d |l-m|} \delta \rho (m) \label{fredholm}, 
\eeq
where $\kp$ is the magnitude of the component of the photon parallel to the 2-d layer, d the separation between the $N+1$ layers, $V(\kp)= \frac{2 \pi e^2}{\kp \varepsilon_M}$ and $\varepsilon_M$ is the material dielectric constant. Here $D_0(\kp,\omega)$ is the "bubble" polarization in 2-d; it is approximated well in terms of the 2-d density $n$ and effective mass $m^*$ by
 $$ D_0\sim \frac{n \kp^2}{m^* \omega^2}. $$
 When  the dielectric constants in the different layers are different, one must also add  image  charges to \disp{fredholm} as  explained in \cite{Jain}, who provide a complete numerical solution for all cases.  

\begin{figure}[htbp]
\centering
\includegraphics[width=\columnwidth]{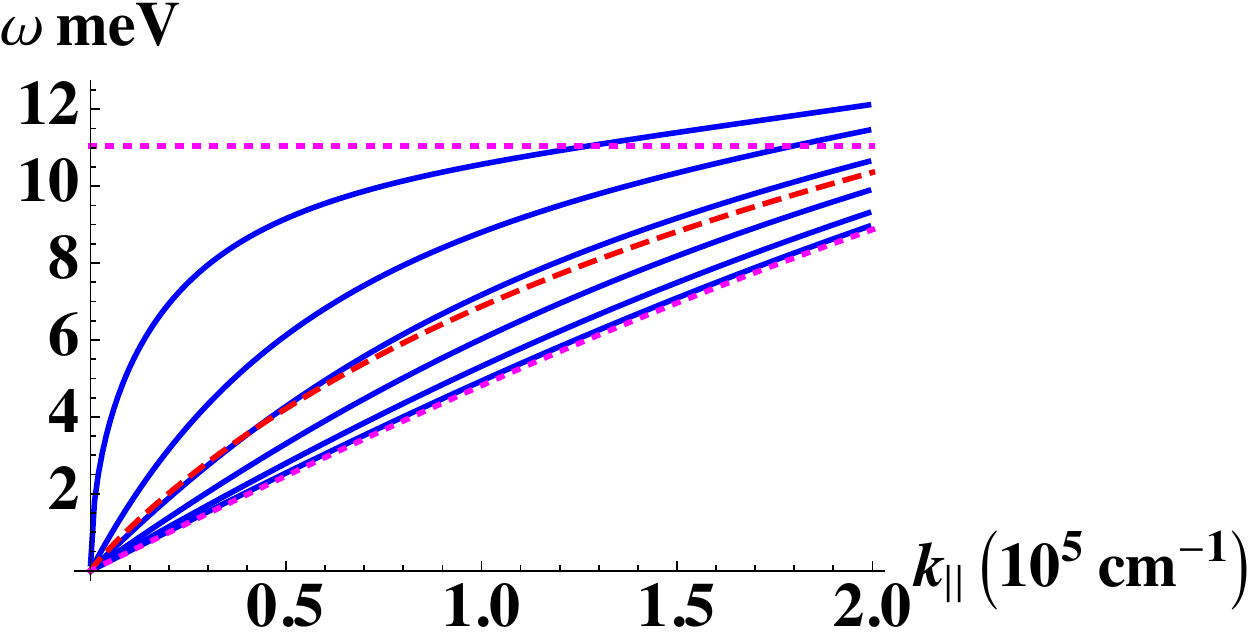}
 \caption{The six plasmon branches for a six layer system in blue solid curves, the geometric mean frequency from \disp{GM} in red dashed curve, and the 3-d bulk and 2-d bulk plasmon in magenta dotted curves.  The parameters used are similar to those of sample 1 in \cite{Pinczuk}, we used d= 900$A^0$, n=7.3$\times10^{11}$ cm$^{-2}$, $m^*=0.07m_e$, $\varepsilon=13.1$.
  \label{Fig1} }
 \end{figure}

Comparing \disp{fredholm} with \disp{Cij} we see that the plasmon frequencies for the $N+1$ layer problem are obtained from $\Lambda_\nu$ in \disp{ev}
\beq
\omega_\nu(\kp) = \sqrt{\frac{2 \pi n e^2}{\varepsilon_M m^*}} \sqrt{\kp \Lambda(q_\nu,\kappa)} \label{plasmon-2}
\eeq
by identifying $\kappa= \kp d$. The allowed $q_\nu$'s are given by \disp{qs}, and are not evenly spaced. The exact determination of the Toeplitz determinant implies that we have a sum-rule on 
\beq
\langle \omega(\kp) \rangle_{gm}&\equiv& \left[\prod_{\nu=0}^N \omega_\nu(\kp)\right]^{\frac{1}{N+1}}\nn \\
&=& \sqrt{\frac{2 \pi n e^2}{\varepsilon_M m^*}} \sqrt{\kp } \left[(1- e^{- 2 \kp d})\right]^{\frac{N}{2 (N+1)}} \label{GM}
\eeq
In \figdisp{Fig1} we illustrate the plasmon solutions for the case of 6 layers using parameters close to those in \cite{Pinczuk}, and also display the geometric mean.

It is useful to note that in general layered systems, the background dielectric function varies between layers , often described as a $\epsilon_0$-$\epsilon$-$\epsilon_0$ configuration of the layers~\cite{Jain}. The pure Toeplitz spectrum is obtained when  $\epsilon_0=\epsilon$. In fact the  experiment in \cite{Pinczuk-2} corresponds to  such a case, with a vanishing dielectric  contrast.

In the generic $\epsilon_0$-$\epsilon$-$\epsilon_0$ configuration of the layers~\cite{Jain}, the problem corresponds to the more complicated Toeplitz-Hankel combination discussed in the previous section, with (in the notation of Ref.~\cite{Jain}, with $N\rightarrow N + 1$) 
\begin{eqnarray} 
	c/a &=& \frac{\epsilon - \epsilon_0}{\epsilon + \epsilon_0} \exp[-\kappa(L - N d)]\nonumber\\
	b/a &=& \left(\frac{\epsilon - \epsilon_0}{\epsilon + \epsilon_0}\right)^2 \exp[-2\kappa (L/d)].
\end{eqnarray}
The right hand side of Eq.(\ref{GM}) is multiplied by 
\begin{equation}
	[(1 - b/a)^{N-1} (1 + c/a)^2]^{1/(N+1)} (1 - b/a)^{-1/2}.
\end{equation}

  \vspace{.25 in}
\noindent {\bf \S Acknowledgments:}

We thank  Professor J. K. Jain and Professor Aaron Pinczuk for a  helpful correspondence, and Professor Torsten Ehrhardt for feedback about prior 
work on Toeplitz matrices. We also thank Professor Peter Forrester for pointing out Ref.\cite{For1}.
The work at UCSC was supported by the US Department of Energy (DOE), Office of Science, Basic Energy Sciences (BES), under Award No. DE-FG02-06ER46319.

 \end{document}